\PassOptionsToPackage{table}{xcolor}
\documentclass[conference]{IEEEtran}
\usepackage[T1]{fontenc}
\usepackage[utf8]{inputenc}
\usepackage{mathptmx}   
\newcommand{\figfont}{\fontfamily{phv}\selectfont}   % Helvetica for Fig. 1 and table badges
\usepackage{graphicx}
\usepackage{booktabs}
\usepackage{amsmath}
\usepackage{siunitx}
\usepackage{tikz}
\usetikzlibrary{arrows.meta,positioning,fit,calc}
\usepackage{xcolor}
\usepackage{hyperref}
\hypersetup{colorlinks=true, urlcolor=blue!50!black, linkcolor=black, citecolor=black}
\usepackage{cite}
\usepackage{enumitem}
\usepackage{makecell}
\usepackage{array}
\usepackage{pifont}
\usepackage{soul}
\usepackage{threeparttable}
\usepackage{multirow}
\newcommand{\cmark}{\ding{51}}\newcommand{\xmark}{\ding{55}}\newcommand{\pmark}{$\circ$}
\setstcolor{red}

\DeclareTextFontCommand{\textbfit}{\bfseries\itshape}

\definecolor{cin}{HTML}{2A78D6}\definecolor{cinf}{HTML}{E8F0FB}
\definecolor{cllm}{HTML}{4A3AA7}\definecolor{cllmf}{HTML}{EEEBFA}
\definecolor{csim}{HTML}{1BAF7A}\definecolor{csimf}{HTML}{E6F6EF}
\definecolor{cref}{HTML}{8A8984}\definecolor{creff}{HTML}{F2F2F0}
\definecolor{cfb}{HTML}{E8590C}
\definecolor{clay}{HTML}{C2338A}\definecolor{clayf}{HTML}{FBEAF3}
\definecolor{cgen}{HTML}{EDA100}\definecolor{cgenf}{HTML}{FDF5DF}
\definecolor{heat}{HTML}{2A78D6}
\definecolor{typeS1}{HTML}{1BAF7A}\definecolor{typeS2}{HTML}{4A3AA7}\definecolor{typeB}{HTML}{2A78D6}
\newcommand{\typebox}[1]{{\setlength{\fboxsep}{1.2pt}\colorbox{type#1}{\textcolor{white}{\bfseries\figfont\tiny #1}}}}
\newcommand{\skillbox}[2]{{\setlength{\fboxsep}{1.4pt}\colorbox{#1}{\textcolor{white}{\bfseries\figfont\tiny #2}}}}
\newcommand{\yes}{\textcolor{csim}{\cmark}}
\newcommand{\prt}{\pmark}
\newcommand{\no}{\textcolor{cref}{\xmark}}
\newcommand{\opera}{OPERA}
\newcommand{\gmid}{$g_m/I_D$}

\begin{document}

\title{\opera{}: OPamp Design by an Expert-Taught Reasoning Agent, from Specifications to Layout}

\author{\IEEEauthorblockN{Vishal Saxena}
\IEEEauthorblockA{\textit{ECE Dept., University of Delaware} \\
Newark, DE, USA \\
vsaxena@udel.edu}
\and
\IEEEauthorblockN{Soumyajit Mandal}
\IEEEauthorblockA{\textit{Instr. Div., Brookhaven National Laboratory} \\
Upton, NY, USA \\
smandal@bnl.gov}
}

\maketitle
\pagestyle{empty}
\thispagestyle{empty}

\begin{abstract}
Analog flows from specifications to a signed-off layout rarely run without manual iteration. Recent large language model
(LLM) agents size circuits with optimizer loops over the device dimensions and stop at the schematic. We present \opera{}
(OPamp design by an Expert-taught Reasoning Agent), in which a general-purpose LLM agent designs CMOS opamps from
specifications to post-layout sign-off with open-source tools. First, the designer teaches the agent the $g_m/I_D$ method:
map the specifications to transconductances and minimum currents, choose $g_m/I_D$ per device role from characterized
lookup tables, and size from equations, leaving a few knobs to search and an explainable design. Second, the agent
learns the gLayout framework and taught layout practice, and writes a reusable layout generator that closes the layout
and extraction loop at designer-approved gates. In SKY130, the agent designed a shared constant-gm bias and six opamps
(5T, folded-cascode and telescopic OTAs; Miller, folded-cascode and telescopic two-stage) for $f_{un}\geq50$~MHz,
PM $\geq60^\circ$, SR $\geq25$~V/$\mu$s into 1~pF and a minimum dc gain of 30 to 60~dB. All six layouts are DRC-clean
and LVS-matched, and their extracted netlists meet every target at all 15 PVT corners ($f_{un}\geq53.7$~MHz,
PM $\geq62.1^\circ$, SR $\geq27.9$~V/$\mu$s). The benchmark netlists are released at \url{https://github.com/AMPICLab/AMPICBench}.
\end{abstract}


\IEEEpeerreviewmaketitle

\section{Introduction}
\IEEEPARstart{A}{n} analog block reaches tape-out only after two nested loops close: the designer sizes,
compensates and re-simulates the circuit until it meets its specifications over process, voltage and
temperature (PVT), then draws, verifies and extracts the layout, and returns to the schematic when parasitics
or layout adjustments move the results. Automating both loops end to end remains a long-standing goal of analog
designers and EDA developers.

LLM-based flows now generate topologies and
netlists~\cite{lai2025analogcoder,gao2025analoggenie,bao2026analogagent}, port amplifiers across
processes~\cite{liu2024ampagent} and assist layout~\cite{liu2024layoutcopilot,hammoud2024glayout,esfandiari2025acdc},
alongside algorithmic layout tools such as ALIGN~\cite{kunal2019align} and MAGICAL~\cite{chen2019magical}.
For sizing, most flows place an optimizer in a loop: the LLM proposes or narrows a search space, and Bayesian
optimization, trust-region search or the LLM itself iterates over the device
dimensions~\cite{yin2024adollm,liu2025eesizer,kochar2024ledro,yu2026autosizer}. These loops spend many
simulations rediscovering what a designer knows from first principles, usually run at the typical corner, and
stop at the schematic.

A designer instead maps the specifications to transconductances, picks the \gmid{} value for each
device from its role, sizes from equations and lookup tables, and tunes a handful of knobs whose effect is
understood. This paper asks whether an LLM agent can be \emph{taught} to design this way, in both the
schematic and the layout loop, and whether the result holds up over PVT after extraction. We focus on single-ended opamps (or OTAs) and a small set of specifications: unity-gain frequency, phase margin, gain, and slew rate into a fixed load. The two-stage opamps are restricted to a Class-A output stage. 
The contributions are:
\begin{enumerate}[leftmargin=*,itemsep=0pt,topsep=2pt]
\item \opera{}, an agent flow from specifications to post-layout PVT sign-off with open-source tools, directed
and taught by us (\textit{i.e.}, the designer) at defined gates (Section~\ref{sec:flow}).
\item Equation-first sizing with per-role \gmid{}, leaving a few knobs instead of tuning each transistor (Section~\ref{sec:sizing}).
\item Agent-learned, designer-taught layout distilled into a reusable generator that closes the layout and
extraction loop (Section~\ref{sec:layout}).
\item A SKY130 case study of six opamps with post-layout PVT results and an optimizer comparison (Section~\ref{sec:case}).
\end{enumerate}

\section{The \opera{} Flow}\label{sec:flow}

\begin{figure*}[!t]
\centering
\includegraphics[width=0.76\textwidth]{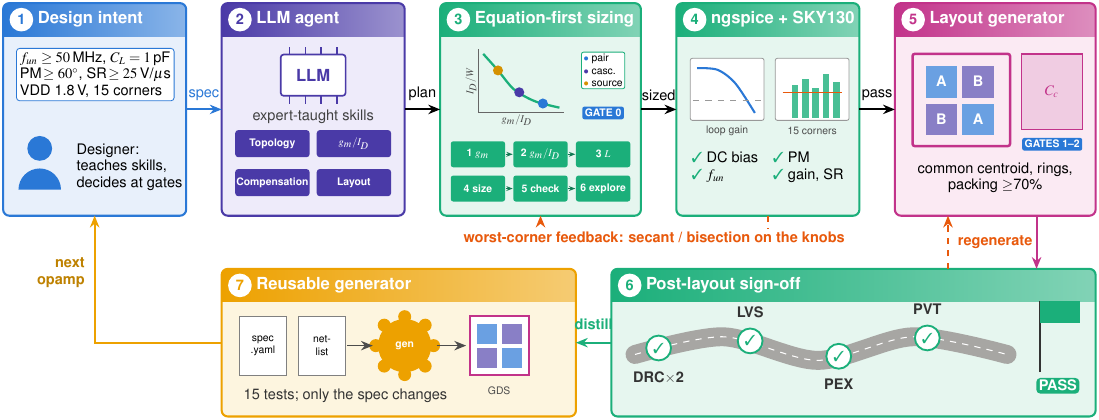}
\caption{The \opera{} flow. (1) The designer gives the specifications, teaches skills and decides at the
gates. (2) A general-purpose LLM agent applies the taught skills and (3) sizes the circuit with the six-step \gmid{} flow (specifications to $g_m$ and minimum
currents, per-role \gmid{}, $L$, sizing from tables, checks, trade-offs), leaving a few knobs. (4) Simulation over 15 PVT corners returns worst-corner
feedback to those knobs. (5) The layout generator builds the cell after the unit and floorplan gates, and
(6) DRC in two decks, LVS, extraction and post-layout PVT simulation sign it off or trigger regeneration.
(7) Each signed-off cell is distilled into a reusable, tested generator that the next opamp reuses.}
\label{fig:flow}
\end{figure*}

\textbf{Fig.~\ref{fig:flow}} depicts our end-to-end flow. The agent is a general-purpose LLM (Claude, Anthropic) in a Linux sandbox with shell, file and simulation tools, without training or fine-tuning. It uses open-source tools, \textit{i.e.}, ngspice~\cite{ngspice}, Magic and Netgen~\cite{magic}, KLayout~\cite{klayout}, SchemDraw \cite{delker2026schemdraw}, the open SKY130
models~\cite{skywater130}, and gLayout~\cite{openfasoc,hammoud2024glayout}. We as the  designer do not edit netlists or layouts; each design starts from one directive, such as ``design a Folded-cascode OTA for 50\,dB gain, $f_{un}\ge 50$~MHz
for a 1\,pF load, and SR$>$25\,$V/\mu s$ for minimum static power consumption.'' The designer's lessons are kept in five skill files (sizing, compensation, gated layout flow, SKY130 layout rules, opamp layout guidelines).

\subsection{Designer-taught skills and decision gates}\label{sec:skills}
 We used Claude's \textit{Opus} 5.5 model for the LLM teaching phase and skill development, then switched to the lower-token-cost \textit{Sonnet} 5.5 model to run the iterative loops. \textbf{Table~\ref{tab:skills}} lists what the designer taught through plain language, lecture slides, short recipes
and worked examples. Each lesson became a skill file of rules and checks that later sessions reload. The designer intervenes only at defined gates: the per-role \gmid{} choices, any sizing
change, any accepted corner failure and, in layout, the unit fingering (gate 1) and the floorplan (gate 2).
At each gate the agent presents two or three simulated options.

\begin{table}[!t]
\centering
\caption{Skills Taught by the Designer}
\label{tab:skills}
\setlength{\tabcolsep}{2.5pt}
\scriptsize
\begin{tabular}{@{}p{14mm}p{33mm}p{32.5mm}@{}}
\toprule
\textbf{Skill} & \textbf{Taught by the designer} & \textbf{Used by the agent as} \\
\midrule
\skillbox{cllm}{Topology} & Signal path, gain and pole nodes, stack headroom, replica bias, folding currents & Headroom budget per stack \\
\skillbox{csim}{$\mathbf{g_m/I_D}$} & Six-step flow: specs to $g_m$ and currents, \gmid{} per role, $L$, size, check, explore & Density, $f_T$, $g_m r_o$ from tables; sizes from equations; worst-corner knob iteration \\
\skillbox{cin}{Compensation} & Two-stage Miller recipe: $m=g_{m2}/g_{m1}$, $C_c$, $R_z$ placement & Starting $C_c$, $R_z$, $m$; one knob per specification \\
\skillbox{clay}{Layout} & Floorplans; common centroid, interdigitation, dummies; rings per stage; packing $\geq$70\%; short parasitics sensitive nodes & Generator templates; matching and packing checkers \\
\bottomrule
\end{tabular}
\end{table}

\section{Equation-First Sizing}\label{sec:sizing}

\subsection{Characterization}
The agent studied the author's lecture slides \cite{saxena2020cpeg624} and the \gmid{} literature~\cite{jespers2017gmid}, then characterized the SKY130
low-$V_{th}$ devices at the planned finger width (3.11~$\mu$m NMOS, 2.48~$\mu$m PMOS), and built lookup tables of current density, $f_T$, $g_m r_o$, and capacitance against \gmid{} for several $L$ at five process corners.

\subsection{The \gmid{} Design Flow}
The designer taught the agent the six steps designers follow~\cite{jespers2017gmid} (\textbf{Fig.~\ref{fig:flow}},
panel 3):
\begin{enumerate}[leftmargin=*,label=\arabic*),itemsep=1pt,topsep=2pt]
\item \textit{Specifications to $g_m$ and currents.} Gain-bandwidth or unity-gain frequency ($f_{un}$) sets $g_{m1}=2\pi\,f_{un}\cdot C_L$ for single-stage, and $g_{m1}=2\pi\,f_{un}\cdot C_c$ for two-stage amplifier; the phase
margin sets $g_{m2}$ through the non-dominant pole, and the slew rate sets the minimum currents. The first
stage slews by steering its tail current into the compensating capacitance, so $I_{tail}\geq\mathrm{SR}\cdot
C_L$ in a single-stage and $I_{tail}\geq\mathrm{SR}\cdot C_c$ in a two-stage amplifier. In a folded-cascode (\textbf{FC}) stage, folding branches must carry at least $I_{tail}$, and a class-A output stage must sink at least $\mathrm{SR}\,(C_L+C_c)$. For 50~MHz
and 25~V/$\mu$s into 1~pF, a single stage needs $g_{m1}\geq314~\mu$S and $I_{tail}\geq25~\mu$A; as
$g_{m1}=(g_m/I_D)_1 I_{tail}/2$, $\mathrm{SR}/f_{un}=4\pi/(g_m/I_D)_1$ caps $(g_m/I_D)_1$ at 25.
\item \textit{\gmid{} per device role.} High for the input pair ($g_m$ per current, noise, offset), low for
mirrors and current sources (for matching, as $\sigma_{\Delta I/I}\propto(g_m/I_D)\,\sigma_{V_T}$, noise and area considerations),
moderate to high for cascodes (headroom), and capped where a device's $f_T$ sets a signal-path pole. Devices that copy a
current share \gmid{} and $L$, so the copy holds over corners. 
\item \textit{Channel length.} $L$ follows from the gain budget ($g_m r_o$ against \gmid{}), traded against
$f_T$ and matching.
\item \textit{Sizing from the tables.} $I_D=g_m/(g_m/I_D)$, or the branch current, and $W=I_D/(I_D/W)$,
rounded to whole fingers. About a dozen continuous widths, lengths and currents reduce to a few knobs: the
\gmid{} of each role, the branch currents and, in a two-stage amplifier, $m=g_{m2}/g_{m1}$, $C_c$ and $R_z$.
\item \textit{Checks and iteration.} Headroom ($V_{D,sat}\approx2/(g_m/I_D)$) and pole estimates come first. The equations using the characterization tables predicted gain and power within 2\% but overestimate the phase margin by
about 18$^\circ$, so stability closes in simulation, one knob per specification at the worst corner, with secant updates for continuous knobs and bisection for integer ones.
\item \textit{Trade-offs.} Design charts (power, gain, swing, $f_T$ against \gmid{}) yield two or three candidates for a gate. 
\end{enumerate}

\subsection{Compensation recipe}\label{sec:comp}
For two-stage amplifiers the agent applies the designer's recipe, based on the two-pole Miller model,
\begin{equation}
\mathrm{PM}\approx 90^\circ-\tan^{-1}\!\left(\frac{C_2}{m\,C_c}\right)-\tan^{-1}\!\left(\frac{1}{m}\right),
\label{eq:pm}
\end{equation}
with $m=g_{m2}/g_{m1}$. With $R_z=1/g_{m2}$ the zero term of (\ref{eq:pm}) vanishes and 60$^\circ$ needs $m\,C_c/C_2\geq1.73$; a larger $R_z$ moves the
zero into the left half plane just above $f_{un}$; the third pole $1/(2\pi R_z C_1)$ must stay well above
$f_{un}$. The agent uses (\ref{eq:pm}) for starting values and knob directions and closes the phase margin in simulation.

\section{Closing the Layout and Extraction Loop}\label{sec:layout}

\subsection{Learning the layout framework}
The agent learned gLayout~\cite{hammoud2024glayout} from its examples and documentation and found its gaps by
running DRC on small pieces: gLayout's layer names were offset by one from SKY130, so the agent wrote an
explicit layer map and routes every net as a Manhattan path with via stacks. It then wrote a row generator for
matched transistor arrays and uses Magic's PDK generators for resistors and MIM capacitors.

\subsection{Taught layout practice and gates}
The designer taught layout from examples and know-how (Table~\ref{tab:skills}). For a two-stage opamp: the
input pair as a two-dimensional common centroid; the load mirror and tail interdigitated
with dummies; each stage in its own guard rings; the Miller capacitor between the second-stage PMOS and NMOS rings, low-impedance plate on the output; and the nulling resistor
between the stages, so high-impedance nodes stay short. At gate 1 the agent proposes unit fingerings and
simulates each over the corners, since finger widths move binned models more than parasitics do; at gate 2 it
compares two or three floorplans by area, utilization and routing cost; gate 3 is the full layout and its
verification. 

\subsection{A reusable layout generator}
The agent then distilled its hand-scripted layouts into a generator for the two-stage family. Its inputs are
the reference netlist and a layout intent file recording the designer's decisions: unit sizes,
matched-set patterns, passive placement and pin edges. One command regenerates the cell and reports
PASS only if Magic and KLayout DRC are clean, Netgen LVS matches, the layout satisfies every item of the intent file, the matching and packing checks pass, and the
extracted netlist meets the design targets over the corners.
Finished blocks (bias, reference, opamp) are then assembled hierarchically: each block exposes its rails and pins as a fixed contract, neighbouring blocks abut and share supply rails, the remaining nets are routed and the flat cell is re-verified with DRC, LVS and extraction.

\begin{table}[!t]
\centering
\caption{Comparison with LLM-Based Analog Design Flows. \yes{} Yes, \prt{} Partly, \no{} No.}
\label{tab:compare}
\setlength{\tabcolsep}{1.3pt}
\scriptsize
\begin{threeparttable}
\begin{tabular}{@{}lccccccccc@{}}
\toprule
\textbf{Work} & \textbf{Topo.} & \textbf{Size}$^b$ & \makecell{$\mathbf{g_m}$/\\$\mathbf{I_D}$} & \textbf{Comp.} & \textbf{PVT} & \textbf{Layout} & \makecell{\textbf{Sign-}\\\textbf{off}$^c$} & \makecell{\textbf{Reus.}\\\textbf{gen.}} & \makecell{\textbf{Taught}\\\textbf{skills}} \\
\midrule
AnalogCoder~\cite{lai2025analogcoder} & \yes & L & \no & \no & \no & \no & \no & \no & \no \\
ADO-LLM~\cite{yin2024adollm} & \no & O & \no & \no & \no & \no & \no & \no & \no \\
LEDRO~\cite{kochar2024ledro} & \no & O & \no & \no & \no & \no & \no & \no & \no \\
AutoSizer~\cite{yu2026autosizer} & \no & O & \no & \no & \no & \no & \no & \no & \no \\
Artisan~\cite{chen2024artisan} & \yes & E & \yes & \prt & \no & \no & \no & \no & \no \\
AmpAgent~\cite{liu2024ampagent} & \prt & E & \no & \prt & \no & \no & \no & \no & \no \\
EEsizer~\cite{liu2025eesizer} & \no & L & \no & \no & \no & \no & \no & \no & \no \\
White-box~\cite{chen2025whitebox} & \no & L & \yes & \prt & \prt & \no & \no & \no & \no \\
AnalogAg.~\cite{bao2026analogagent} & \prt & O & \no & \no & \no & \no & \no & \no & \no \\
LayoutCop.~\cite{liu2024layoutcopilot} & \no & \no & \no & \no & \no & \yes & \prt & \no & \no \\
Glayout NL~\cite{hammoud2024glayout} & \no & \no & \no & \no & \no & \yes & \prt & \prt & \no \\
ACDC~\cite{esfandiari2025acdc} & \prt & \no & \no & \no & \no & \no & \no & \no & \no \\
\midrule
\rowcolor{cllmf}\textbf{\opera{} (ours)} & \prt$^a$ & \textbf{E} & \yes & \yes & \yes & \yes & \yes & \yes & \yes \\
\bottomrule
\end{tabular}
\begin{tablenotes}\footnotesize
\item From published descriptions. $^a$Topologies chosen by the designer and taught to the agent. \prt{} in Topo.: from literature or code generation.
$^b$O: optimizer loop; L: LLM-driven iteration; E: equations, then a few knobs. $^c$DRC, LVS and extraction
over PVT.
\end{tablenotes}
\end{threeparttable}
\end{table}

\section{Case Study}\label{sec:case}

\subsection{Specifications and shared bias}
All amplifiers target open-loop $f_{un}\geq50$~MHz with $C_L=1$~pF in a loop-gain analysis ~\cite{tian2001stb} testbench similar to Spectre's \textit{stb} analysis, PM $\geq60^\circ$, SR
$\geq25$~V/$\mu$s for a 0.7-to-1.1-V follower step (20--80\%), and a minimum DC gain $A_0$ of 30~dB (5T), 50~dB (FC and telescopic OTAs) or 60~dB (two-stage opamps), VDD = 1.8~V and $V_{CM}=0.9$~V (testbenches and netlists released as a benchmark~\cite{ampicbench}), biased on chip
by a constant-$g_m$ beta-multiplier reference (BMR) feeding a wide-swing cascode
bias~\cite{baker2019cmos,vsaxena_indirect2006}. As the BMR transconductance tracks $1/R$, the agent chose an
n+ diffusion resistor (+1435~ppm/K) over high-resistance poly. The short-channel BMR \cite{saxena2020cpeg624} provides a 20.0~$\mu$A reference
at tt (19.1 to 21.3 over corners), with loop PM 72.4 to 74.6$^\circ$.

\begin{figure*}[!t]
\centering
\begin{minipage}[b]{0.45\textwidth}
  \centering
  \includegraphics[width=\linewidth]{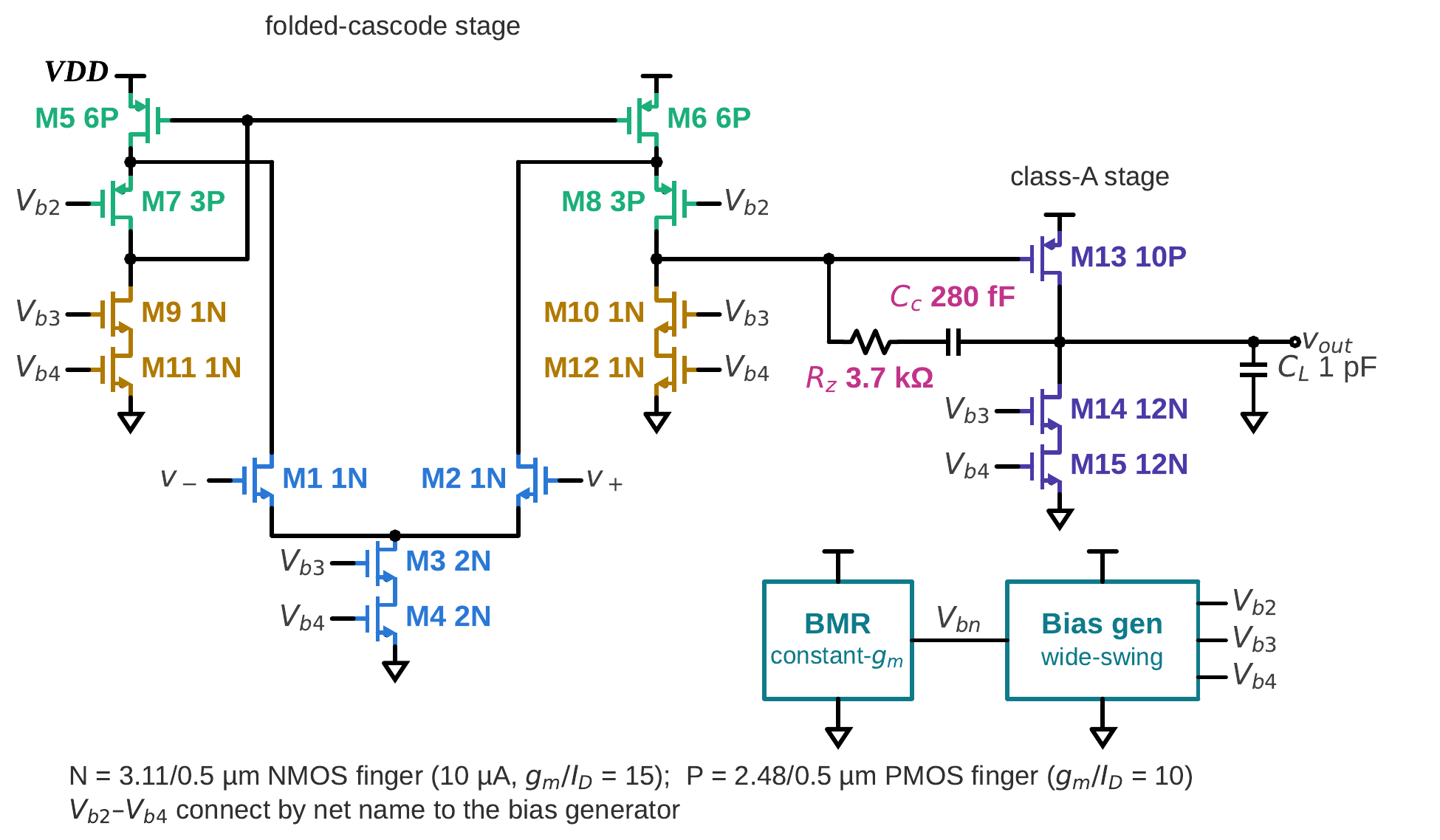}\\[-1pt]
  {\figfont\footnotesize (a)}
\end{minipage}\hspace{0.05\textwidth}%
\begin{minipage}[b]{0.45\textwidth}
  \centering
  \includegraphics[width=\linewidth]{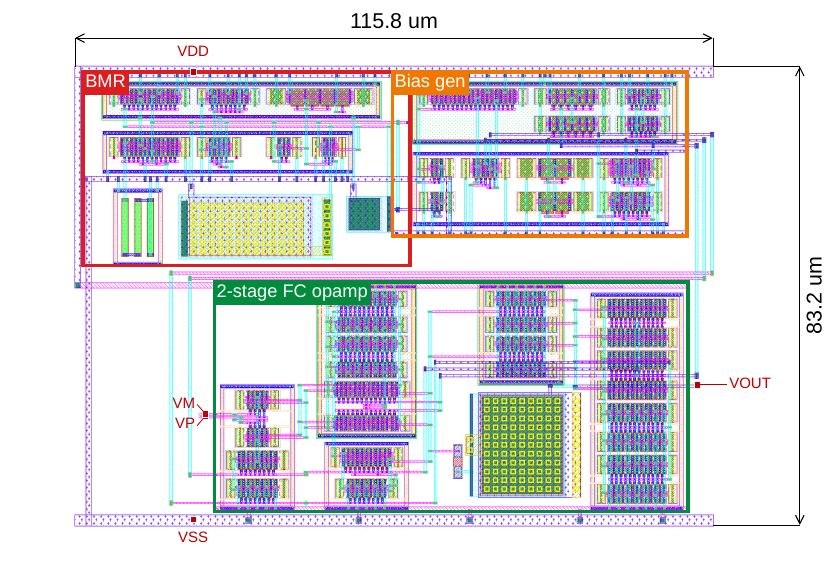}\\[-1pt]
  {\figfont\footnotesize (b)}
\end{minipage}
\caption{Two-stage opamp with a folded-cascode first stage and a class-A second stage, designed by \opera{}.
(a) Schematic drawn by the agent from its sized netlist, devices colored by role; labels are finger counts. (b) Its automatically generated layout; DRC/LVS clean and meets specs in all corners.}
\label{fig:opamp}
\end{figure*}

\subsection{Schematic results}
The agent designed six topologies from the designer's Cadence starting points in SKY130 PDK: 5T, folded-cascode (FC) and telescopic OTAs (the latter with a replica half-circuit for its cascode bias), a Miller opamp, and an FC or telescopic stage followed by a class-A stage. Fig.~\ref{fig:opamp}(a) shows the FC two-stage.
Table~\ref{tab:results} lists the knobs and the results over 15 corners. Every design meets $f_{un}\geq50$~MHz, PM
$\geq60^\circ$, SR $\geq25$~V/$\mu$s and its $A_0$ target at every corner. The lowest slew rates, 27.1 and 28.6~V/$\mu$s, are in the
two-stage opamps with cascode first stages, set by the first-stage current charging $C_c$. One limit remains: at fs, 70\,$^\circ$C the weakest cascode is 23 to 63~mV into triode in five designs, yet all targets are met.

\begin{table}[!t]
\centering
\caption{Opamp Results over 15 PVT Corners (Schematic, $C_L$=1~pF, VDD=1.8~V, $V_{CM}$=0.9~V, On-Chip Bias)}
\label{tab:results}
\setlength{\tabcolsep}{2.2pt}
\scriptsize
\begin{threeparttable}
\begin{tabular}{@{}cllcccccc@{}}
\toprule
 & \textbf{Circuit} & \textbf{Knobs}$^a$ & \makecell{$\mathbf{A_0}$\\(dB)} & \makecell{$\mathbf{f_{un}}$\\(MHz)} & \makecell{\textbf{PM}\\($^\circ$)} & \makecell{\textbf{SR}$^b$\\(V/$\mu$s)} & \makecell{\textbf{I$_{\text{DD}}$}\\($\mu$A)} \\
\midrule
\typebox{S1} & 5T OTA & $n$ & 31.1--31.8 & \cellcolor{heat!46}68.1--76.0 & \cellcolor{heat!57}82.4--84.1 & \cellcolor{heat!31}35.4 & 126--144 \\
\typebox{S1} & FC OTA & $n_1,n_2$ & 51.1--53.0 & \cellcolor{heat!43}66.5--77.4 & \cellcolor{heat!32}67.6--72.6 & \cellcolor{heat!27}33.2 & 282--358 \\
\typebox{S1} & Telescopic & $n$ & 51.9--55.0 & \cellcolor{heat!70}96.6--115.2 & \cellcolor{heat!38}71.2--77.9 & \cellcolor{heat!41}41.4 & 243--307 \\
\midrule
\typebox{S2} & Miller & $n,m,C_c,R_z$ & 60.8--64.0 & \cellcolor{heat!70}82.2--100.8 & \cellcolor{heat!35}69.1--75.7 & \cellcolor{heat!50}46.1 & 338--422 \\
\typebox{S2} & FC 2-stage & $n,m,C_c,R_z$ & 63.2--68.2 & \cellcolor{heat!25}57.5--72.8 & \cellcolor{heat!36}70.2--78.9 & \cellcolor{heat!18}28.6 & 330--424 \\
\typebox{S2} & Tele.\ 2-stage & $n,m,C_c,R_z$ & 68.7--78.0 & \cellcolor{heat!61}75.5--95.8 & \cellcolor{heat!30}66.8--75.6 & \cellcolor{heat!14}27.1 & 330--423 \\
\bottomrule
\end{tabular}
\begin{tablenotes}\footnotesize
\item \typebox{S1} single-stage; \typebox{S2} two-stage, class-A output. $^a$$n$: pair fingers, $n_1$/$n_2$:
pair/folding fingers. $^b$Worst edge, 0.7--1.1~V step, 20--80\%. Shading: worst-corner margin. I$_{\text{DD}}$ includes the bias. Corners: tt,
ff, ss, fs, sf at 0/27/70\,$^\circ$C. $A_0$ targets: 30, 50, 50, 60, 60, 60~dB.
\end{tablenotes}
\end{threeparttable}
\end{table}

\subsection{Comparison with Optimizer Search}\label{sec:baseline}
AutoSizer~\cite{yu2026autosizer} draws its optimizer from a pool that includes Bayesian optimization and simulates only
the typical corner. To see how such optimizers fare over PVT, random search and the tree-structured Parzen estimator
(TPE)~\cite{bergstra2011tpe}, a Bayesian method, in
Optuna~\cite{akiba2019optuna}, sized the same six circuits with the same bias, testbenches and 15 corners (3 seeds, 5 for the 5T; 500-run cap). It uses the first-campaign spec (no gain target).
The cost is ngspice runs, a candidate stops at its first failing corner, and a pass means all 15 corners; \opera{} is
counted the same way on its stored sweeps. Protocol A is the specification plus the \opera{} I$_{\text{DD}}$; B adds a worst saturation margin
$\geq-50$~mV (5T: $\geq0$). Under A, TPE passes every circuit in 45 to 204 runs (median) against 30 (54 for the 5T) for \opera{}, but with margins down to
$-169$~mV. Under B it passes 5 of 15 runs on the five larger circuits and never on the FC and telescopic two-stage designs, where \opera{} needs
55 (Table~\ref{tab:base}).

\begin{table}[!t]
\centering
\caption{Runs to First Passing Design: Passing Seeds (Median Runs)}
\label{tab:base}
\setlength{\tabcolsep}{2.4pt}
\scriptsize
\begin{threeparttable}
\begin{tabular}{@{}cl c cc cc@{}}
\toprule
 & & \textbf{\opera{}}$^a$ & \multicolumn{2}{c}{\textbf{Protocol A (spec)}} & \multicolumn{2}{c}{\textbf{Protocol B (sat.)}} \\
\cmidrule(lr){4-5}\cmidrule(lr){6-7}
 & \textbf{Circuit} & A / B & Random & TPE & Random & TPE \\
\midrule
\typebox{S1} & 5T OTA & 54 / 54 & \cellcolor{heat!70}5/5 (58) & \cellcolor{heat!70}5/5 (48) & \cellcolor{heat!70}5/5 (76) & \cellcolor{heat!70}5/5 (48) \\
\typebox{S1} & FC OTA & 30 / 59 & \cellcolor{heat!70}3/3 (250) & \cellcolor{heat!70}3/3 (129) & \cellcolor{heat!23}1/3 (368) & \cellcolor{heat!23}1/3 (32) \\
\typebox{S1} & Telescopic & 30 / 59 & \cellcolor{heat!70}3/3 (35) & \cellcolor{heat!70}3/3 (45) & \cellcolor{heat!47}2/3 (258) & \cellcolor{heat!47}2/3 (41) \\
\midrule
\typebox{S2} & Miller & 30 / 53 & \cellcolor{heat!70}3/3 (40) & \cellcolor{heat!70}3/3 (204) & \cellcolor{heat!70}3/3 (40) & \cellcolor{heat!47}2/3 (181) \\
\typebox{S2} & FC 2-stage & 30 / 55 & \cellcolor{heat!47}2/3 (281) & \cellcolor{heat!70}3/3 (141) & 0/3 & 0/3 \\
\typebox{S2} & Tele.\ 2-stage & 30 / 55 & \cellcolor{heat!70}3/3 (217) & \cellcolor{heat!70}3/3 (166) & \cellcolor{heat!23}1/3 (352) & 0/3 \\
\bottomrule
\end{tabular}
\begin{tablenotes}\footnotesize
\item $^a$Runs under protocol A / B. Shading: pass rate.
\end{tablenotes}
\end{threeparttable}
\end{table}

\subsection{Layout results}
Fig.~\ref{fig:opamp}(b) shows the FC two-stage layout, and Table~\ref{tab:layout} summarizes the six layouts and the bias
cell. All six are DRC clean in both decks and LVS matched, and their extracted netlists meet every target at all 15 corners.
Extraction lowers $f_{un}$ by 2 to 7\% and PM by 1 to 8$^\circ$, and moves $A_0$ by 0.1~dB at most. Three first extractions missed the PM (FC 57.2, telescopic 53.3, FC two-stage 59.4$^\circ$); shorter source $L$ (FC, 1.0~$\mu$m), shorter cascode $L$ (telescopic, 0.6~$\mu$m) and $R_z=3.7$~k$\Omega$ (FC two-stage) fixed them. Layout area utilization is 57 to 89\%. Telescopic layouts hold the bias replica in the same arrays as the devices it copies; pairs, sources and cascodes are centroided, single-unit devices are not.

\begin{table}[!t]
\centering
\caption{Layouts and Post-Layout Results (SKY130, Extracted R and C, Ideal $R_z$ and $C_c$, 15 Corners)}
\label{tab:layout}
\setlength{\tabcolsep}{1.9pt}
\scriptsize
\begin{threeparttable}
\begin{tabular}{@{}clcccccccc@{}}
\toprule
 & \textbf{Cell} & \makecell{\textbf{Area}\\($\mu$m$^2$)} & \textbf{Util.} & \makecell{\textbf{DRC}$^a$\\\textbf{LVS}} & \makecell{$\mathbf{A_0}$ (dB)\\PEX} & \makecell{\textbf{PM} ($^\circ$)\\PEX} & \makecell{$\mathbf{f_{un}}$ (MHz)\\PEX} & \makecell{\textbf{SR}\\(V/$\mu$s)} \\
\midrule
\typebox{S1} & 5T OTA & 496 & \cellcolor{heat!59}89.2\% & 0/0 \yes & \cellcolor{heat!17}31.1--31.8 & \cellcolor{heat!56}81.8--83.5 & \cellcolor{heat!43}66.5--74.3 & \cellcolor{heat!29}34.3 \\
\typebox{S1} & FC OTA & 3482 & \cellcolor{heat!39}69.0\% & 0/0 \yes & \cellcolor{heat!16}51.0--53.0 & \cellcolor{heat!25}63.5--68.6 & \cellcolor{heat!37}63.5--74.0 & \cellcolor{heat!24}31.5 \\
\typebox{S1} & Telescopic & 3015 & \cellcolor{heat!57}86.8\% & 0/0 \yes & \cellcolor{heat!21}51.9--54.9 & \cellcolor{heat!30}66.2--72.9 & \cellcolor{heat!70}92.4--109.9 & \cellcolor{heat!41}41.4 \\
\midrule
\typebox{S2} & Miller & 2356 & \cellcolor{heat!48}78.3\% & 0/0 \yes & \cellcolor{heat!15}60.8--64.0 & \cellcolor{heat!27}64.6--71.0 & \cellcolor{heat!67}78.8--95.6 & \cellcolor{heat!50}46.1 \\
\typebox{S2} & FC 2-stage$^c$ & 3570 & \cellcolor{heat!37}66.9\% & 0/0 \yes & \cellcolor{heat!29}63.2--68.3 & \cellcolor{heat!23}62.1--70.7 & \cellcolor{heat!17}53.7--68.0 & \cellcolor{heat!17}27.9 \\
\typebox{S2} & Tele.\ 2-stage & 2893 & \cellcolor{heat!27}57.3\% & 0/0 \yes & \cellcolor{heat!61}68.5--78.0 & \cellcolor{heat!24}63.1--72.7 & \cellcolor{heat!52}71.1--89.5 & \cellcolor{heat!21}30.5 \\
\midrule
\typebox{B}  & BMR             & 1092 & \cellcolor{heat!47}76.5\% & 0/0 \yes & --- & 61.2--65.5$^b$ & 52.8--81.2$^b$ & --- \\
\bottomrule
\end{tabular}
\begin{tablenotes}\footnotesize
\item \typebox{B} bias cell. $^c$Area and utilization of Fig.~\ref{fig:opamp}(b); PEX is from the final layout (2688~$\mu$m$^2$, 59.7\%). Ranges: 5 process corners at 0/27/70\,$^\circ$C; SR: worst edge.
\end{tablenotes}
\end{threeparttable}
\end{table}

\section{Comparison with Prior Work}\label{sec:discussion}
Table~\ref{tab:compare} positions this work. LEDRO~\cite{kochar2024ledro} and AutoSizer~\cite{yu2026autosizer}
use the LLM to steer an optimizer over the device dimensions; Artisan~\cite{chen2024artisan}
and~\cite{chen2025whitebox} give the LLM \gmid{} data, but it still chooses each width. In \opera{} the designer teaches the \gmid{} flow, so the agent sizes from equations and searches a few knobs. Layout
agents~\cite{liu2024layoutcopilot,hammoud2024glayout,esfandiari2025acdc} do not close the loop through extraction and
PVT; \opera{} does, and leaves a reusable generator. Its scope, one designer and one generator family, keeps results traceable, and extending to more designers and families is the next step.

\section{Conclusion}
\opera{} shows that a general-purpose LLM agent, taught the \gmid{} flow and layout practice by a designer,
can take CMOS opamps from specifications to post-layout sign-off with open-source tools. In SKY130, six opamps
on a shared constant-gm bias meet their dc-gain targets, $f_{un}\geq50$~MHz, SR $\geq25$~V/$\mu$s and
$60^\circ$ of phase margin (for two-stage) at all 15 PVT corners, in schematic and in extracted layouts. The benchmark circuit
netlists are available at \cite{ampicbench}. 

\bibliographystyle{IEEEtran}
\bibliography{bibtex/refs}

\end{document}